# Isospin Diffusion Observables in heavy ion reactions

T.X. Liu[1], W.G. Lynch, M.B. Tsang, X.D. Liu, R. Shomin, W.P. Tan, G. Verde, A. Wagner[2], H.F. Xi[3], H.S. Xu[4],

*National Superconducting Cyclotron Laboratory and Department of Physics and Astronomy, Michigan State University, East Lansing, MI 48824, USA,*

B. Davin, Y. Larochelle, R. T. de Souza,

*Department of Chemistry and IUCF, Indiana University, Bloomington, IN 47405, USA,*

R.J. Charity, and L.G. Sobotka,

*Department of Chemistry, Washington University, St. Louis, MO 63130, USA*

**Abstract**

Collisions of $^{112}$Sn and $^{124}$Sn nuclei, which differ in their isospin asymmetry, provide information about the rate of isospin diffusion and equilibration. While several different probes can provide accurate diffusion measurements, the ratios of the mirror nuclei may be the simplest and most promising one. Ratios of the mass seven mirror nuclei yields are analyzed to show the rapidity, transverse momentum and impact parameter dependence of isospin diffusion.

---

[1] Present address: University of California, Davis.
[2] Present address: Institut für Strahlenphysik, Forschungszentrum Rossendorf, D-01314 Dresden, Germany.
[3] Present address: Benton Associate, Toronto, Ontario, Canada.
[4] Present address: Institute of Modern Physics, Lanzhou, China

Many investigations have provided theoretical guidance and experimental constraints on the equation of state of symmetric nuclear matter [1]. Dense macroscopic nuclear systems like neutron stars, however, are neutron rich and require for their description an understanding of the density dependence of the symmetry energy term in the nuclear Equation of State [2]. Few constraints on this density dependence exist [3], prompting the development of new techniques [4-7] for its determination. Recently, constraints on the density dependence of the symmetry energy were obtained from measurements of isospin diffusion in peripheral nuclear collisions [6,8]. In this paper, we identify a set of experimental observables, specifically observables constructed with yield ratios of mirror nuclei, that provide consistent measures of the isospin diffusion and extend those experimental investigations to a wider range of rapidity, transverse momentum and impact parameter.

In a heavy ion collision involving a projectile and a target with different proton fractions, Z/A, the symmetry energy tends to propel the system towards isospin equilibrium so that the difference between neutron and proton densities is minimized [7]. The isospin asymmetry $\delta = \frac{N-Z}{A}$ of a projectile-like residue produced in a peripheral collision reflects the exchange of nucleons with the target; significant diffusion rates should lead to larger isospin asymmetries for collisions with neutron-rich targets and smaller isospin asymmetries for collisions with proton-rich targets [6].

To isolate the isospin diffusion effects from similar effects caused by pre-equilibrium emission, Coulomb or sequential decays, relative comparisons involving different targets are important. In recent studies, isospin diffusion has been measured by "comparing" A+B collisions of a neutron-rich (A) nucleus and a proton-rich (B) nucleus to symmetric collisions involving two neutron-rich nuclei (A+A) and two proton-rich (B+B) nuclei under the same experimental conditions [6]. Non-isospin diffusion effects such as preequilibrium emission from a neutron-rich (A) projectile should be approximately the same for asymmetric A+B collisions as for symmetric A+A collisions.

Similarly, non-isospin diffusion effects from a proton-rich (B) projectile in B+A collisions and B+B collisions should be the same.

The degree of isospin equilibration can be quantified by rescaling the isospin asymmetry δ of a projectile-like residue from a specific collision according to the isospin transport ratio $R_i(\delta)$ [6,9] given by

$$R_i(\delta) = 2\frac{\delta - (\delta_{A+A} + \delta_{B+B})/2}{\delta_{A+A} - \delta_{B+B}}. \tag{1}$$

In the absence of isospin diffusion, the asymmetry $\delta_{A+B}$ of a residue of a neutron-rich projectile following a collision with a proton-rich target has the limiting values $R_i(\delta_{A+B}) = R_i(\delta_{A+A}) = 1$. Likewise without diffusion, $R_i(\delta_{B+B}) = R_i(\delta_{B+A}) = -1$. On the other hand, if isospin equilibrum is achieved for roughly equal sized projectile and target nuclei, $R_i(\delta_{A+B}) = R_i(\delta_{B+A}) \approx 0$. By focusing on the differences in isospin observables between mixed and symmetric systems, $R_i(\delta)$ largely removes the sensitivity to preequilibrium emission and enhances the sensitivity to isospin diffusion.

Ideally, one would like to know the asymmetry of the projectile-like residue immediately after the collision and prior to secondary decay because this is the quantity that is calculated in transport theory [6]. To do this, one can measure an observable X that is linearly dependent on the residue asymmetry, i.e. X = a·δ+b, and constructs the corresponding isospin transport ratio $R_i(X)$

$$R_i(X) = 2\frac{X - (X_{A+A} + X_{B+B})/2}{X_{A+A} - X_{B+B}}.$$

Then, trivial substitution provides that $R_i(X)=R_i(\delta)$ and one dispenses with most of the uncertainty associated with determining δ from measurements of X.

The above idea has been adopted to study the isospin diffusion using the isoscaling observable (X=α) involving two asymmetric collisions $^{124}$Sn+$^{112}$Sn (A+B) and $^{112}$Sn+$^{124}$Sn (B+A) and two symmetric collisions $^{124}$Sn+$^{124}$Sn (A+A) and $^{112}$Sn+$^{112}$Sn (B+B) [6]. The isoscaling parameter α can be obtained from the isotope yield ratios from two reactions, which are similar in all aspects except in their isospin compositions:

$$R_{21}(N,Z)=Y_2(N,Z)/Y_1(N,Z)= C\exp(\alpha N+\beta Z) \qquad (2)$$

where $Y_i(N,Z)$ is the yield of the measured fragments with neutron number N and proton number Z emitted in reaction i (i=1,2), $\alpha$, $\beta$ and C are obtained by fitting the isotope yield ratios to Eq. (2); $\alpha$ is the neutron isoscaling factor, $\beta$ is the proton isoscaling factor and C is the normalization constant.

An expression for the dependence of $\alpha$ for particles evaporated from an excited nucleus of asymmetry $\delta$ have been derived in ref. [10] and an identical expression has been obtained for an equilibrium multifragment decay in ref. [11]. Using the logic of refs. [10,11], one obtains the following expressions for $\alpha$ and $\beta$ that may be applicable to both evaporative and multifragment decays of an excited projectile-like fragment:

$$\alpha = \Delta\mu_n/T = 2C_{sym}(\Delta\delta)(1-\bar{\delta})/T,$$
$$\beta = \Delta\mu_p/T = -2C_{sym}(\Delta\delta)(1+\bar{\delta})/T, \qquad (3)$$
$$\text{and}$$
$$\alpha - \beta = (\Delta\mu_n - \Delta\mu_p)/T = 4C_{sym}(\Delta\delta)/T,$$

where, $\Delta\delta=\delta_2-\delta_1$ and $\bar{\delta}=(\delta_2+\delta_1)/2$ are the differences and mean of the asymmetries of the emitting source, $C_{sym}$ is the coefficient of the symmetry energy term in the nuclear Gibbs free energy. Equations 3 implies that $\alpha-\beta$ depends linearly on the asymmetry $\delta$ of the systems from which the fragments are emitted. For most systems $\delta \ll 1$, making the dependence of $\alpha$ and $\beta$ on $\delta$ essentially linear as well. Both statistical [10,11] and dynamical calculations [12] are consistent with a linear dependence of $\alpha$, $\beta$, and $\alpha-\beta$ on $\delta$.

More convincing would be to demonstrate experimentally the linear dependence between $\alpha$, $\beta$, and $\alpha-\beta$ on $\delta$. In the upper panel of Figure 1, we compare the values for $\alpha$, $\beta$, and $\alpha-\beta$ obtained at mid-rapidities for central $^{112}$Sn+$^{112}$Sn ($\delta = 0.107$), $^{112}$Sn+$^{124}$Sn ($\delta = 0.153$) and $^{124}$Sn+$^{124}$Sn ($\delta = 0.194$) collisions as a function of the total asymmetry using the data of ref. [13]. By construction, $\alpha=\beta=\alpha-\beta=0$ for the reference $^{112}$Sn+$^{112}$Sn system. To ensure complete mixing of projectile and target nucleons, the data in Figure 1 were obtained at center of mass angles of $70°\leq \theta_{CM} \leq 110°$. This avoids the possible contributions of projectile nucleons near $\theta_{CM}=0°$ and target nucleons near $\theta_{CM}=180°$ that are not strongly scattered by the collision. Consistent with theoretical predictions, the measured trends for $\alpha$, $\beta$, and $\alpha$-$\beta$ are linear as seen by the data points joined by the solid, dot-dashed and dashed lines respectively.

The quantity $\alpha$-$\beta$ is of particular interest as it arises naturally from the ratios of yields of mirror nuclei Y(N,Z) and Y(Z,N), where |N-Z|=1.

$$r_A = Y_2(N,Z)/Y_1(Z,N) = C\exp(\alpha-\beta) \qquad (6)$$

or

$$X_A = \ln(r_A) = \alpha-\beta + \ln(C) \qquad (7)$$

where A=N+Z and C is a constant. By using mirror nuclei, one avoids mass dependences in the production mechanisms in the kinematics and in the effect of sequential decays. The latter is minimized in the ratio because both nuclei have similar analogue states. The three most widely used pairs of mirror nuclei are ($^3$H, $^3$He); ($^7$Li, $^7$Be) and ($^{11}$B, $^{11}$C). Values of $X_A$ obtained for A=7 and 11 from Eq. 7 using the published data of ref. [13] are shown in the lower panel of Figure 1. Both $X_7$ and $X_{11}$ depend linearly on $\delta$.

For the subsequent analyses shown below, we choose $X_7$ because the yields of $^7$Li and $^7$Be are measured with reasonably high statistics. We use $X_7$ to explore how the isospin asymmetry varies with rapidity, impact parameter, and transverse momentum. Such explorations require the ability to extract information about the isospin asymmetry from small regions of phase space. If one does not employ $X_7$, this could be a difficult task because the Coulomb force, thermal and collective motion influence the momentum distributions of various fragments differently. While isotopes of an element experience the same Coulomb momentum transfers, their collective momentum contributions are proportional to mass, and their thermal momentum contributions are proportional to the square root of their masses. If an isotopic effect is observed, one would like to be certain that it is due to isospin diffusion and not a kinematic effect. In this paper, we will use the isospin transport ratio constructed with the observable $X_7$ to demonstrate the cancellation of the Coulomb effect leading to the measurement of isospin diffusion as a function of impact parameter, rapidity and transverse momentum.

The isotopic distributions produced in $^{112}$Sn+$^{112}$Sn, $^{112}$Sn+$^{124}$Sn, $^{124}$Sn+$^{112}$Sn and $^{124}$Sn+$^{124}$Sn collisions were measured at the National Superconducting Cyclotron Laboratory at Michigan State University by bombarding $^{112}$Sn and $^{124}$Sn targets of 5

mg/cm$^2$ areal density with 50 MeV per nucleon $^{112}$Sn and $^{124}$Sn beams. Some aspects of these reactions have been published previously [6,13-15]. In the experiment, charged particles were measured with two detection arrays. The multiplicity array consists of 188 plastic scintillator - CsI(Tl) phoswich detectors of the Miniball/Miniwall array [16]. This array provided isotopic resolution for H and He nuclei and elemental resolution for intermediate mass fragments (IMF) with $3 \leq Z \leq 20$.

In addition to the multiplicity array, the projectile-like residues are detected with a Ring Counter Forward Array consists of a double-sided 280 μm thick annular strip Silicon detector backed by an annular array of sixteen 2 cm thick CsI(Tl) detectors. The annular Si detector is segmented into four quadrants with sixteen 1.5 mm wide annular strips arranged in circles of increasing radius extending radially from the inner edge of the detector (48 mm diameter) to the outer edge of the detector (96 mm diameter). This provided resolution in the polar angle. The asymuthal angles of the silicon detector were subdivided into 16 azimuthal pads on the opposite side of the detector. The Ring Counter Forward array is centered around the beam axis and covers the polar angles from 2.2° to 4.5° deg. It provides element resolution for fragments with $3 \leq Z \leq 55$.

The third array, called the Large Angle Si Strip Array (LASSA) [17,18], consists of 9 telescopes each of which comprises of one 65 μm Si, one 500 μm Si and four 6-cm thick CsI detectors. The 50mm x 50mm lateral dimensions of each telescope are divided into 256 square pixels ($\approx$3x3 mm$^2$), providing an angular resolution of about ±0.43°. The Si detectors are backed by four 6 cm long CsI detectors. The center of the device was located at a polar angle of θ=32° with respect to the beam axis, covering polar angles of $7° \leq \theta \leq 58°$. The array provided isotopic resolution Z=1-8. This array provides excellent angular coverage for particles emitted in both peripheral and central collisions. Impact parameters were selected by assuming the multiplicity of charged particles, measured with LASSA [17,18] and the Miniball/Miniwall array [16] decrease monotonically with impact parameter; the combined apparatus covered 80% of the total solid angle.

Figure 2 shows the measured correlation between the charge of the projectile-like residue detected in the ring counter and the observed total charged particle multiplicity (measured in the miniball/miniwall and LASSA arrays). The projectile-like residue decreases monotonically in charge with decreasing impact parameter and appears to vanish at $b/b_{max} \leq 0.4$. There is no evidence supporting the existence of a projectile-like residue in central collisions defined to be $b/b_{max} < 0.2$. Isotopic distributions for midrapidity particles, which originate predominantly from the disintegration of the participant zone were measured with the gates $b/b_{max} < 0.2$ and $70° \leq \theta_{CM} \leq 110°$ and published in Ref. [19].

Figure 3 shows the isotopic distributions from peripheral collisions measured with impact parameter gate $b/b_{max} > 0.8$ and projectile rapidity, $y/y_{beam} > 0.8$ [6]. The solid lines are drawn through the data for the $^{124}$Sn+$^{124}$Sn reaction. The other lines in the figure are extrapolated from the data measured for the $^{124}$Sn+$^{124}$Sn reaction using Eq. 1 and the isoscaling parameters in Ref. [6]. These data and their isoscaling parameters were used in Ref. [6] to investigate the isospin diffusion. Values of $R_i(\alpha) = 0.47 \pm 0.05$ corresponding to the decay of the $^{124}$Sn projectile from the $^{124}$Sn+$^{112}$Sn reaction and $R_i(\alpha) = -0.45 \pm 0.05$ corresponding to the decay of the $^{112}$Sn projectile from the $^{112}$Sn+$^{124}$Sn reaction were obtained. These are shown by the shaded rectangles in Figure 3; the $^{124}$Sn projectile data are shown at $0.7 < y/y_{beam} < 1.0$ and $^{112}$Sn projectile data are shown at $0 < y/y_{beam} < 0.3$ as the latter correspond to the target in $^{124}$Sn+$^{112}$Sn reaction.

Now, we turn to the extraction of isospin transport ratios for the mirror nuclei. By combining the asymmetric collisions of $^{112}$Sn+$^{124}$Sn and $^{124}$Sn+$^{112}$Sn, we obtain complete coverage from projectile to target rapidity. The top two panels of Figure 5 shows the background and efficiency corrected $\frac{dM}{dy d(p_\perp/A)}$ differential multiplicity distributions for $^7$Li and $^7$Be fragments for the $^{124}$Sn+$^{112}$Sn reaction as a function of normalized rapidity $y/y_{beam}$ and transverse momentum $p_t/A$. Here, both the particle rapidity and the beam rapidity are given in the laboratory frame. There is an angular cutoff in the experimental data at $\theta_{lab} < 7°$, corresponding to a rapidity dependent cut on the transverse momentum $p_\perp$. The projectile-like rapidity region near $y/y_{beam} \sim 1$ is primarily obtained from the reaction $^{124}$Sn+$^{112}$Sn using $^{124}$Sn as the projectile while the target-like rapidity region near $y/y_{beam} \sim 0$ was transformed from the data of the inverse reaction, $^{112}$Sn+$^{124}$Sn. Since $^7$Li is a neutron rich isotope, its yield is larger near the original rapidity ($y/y_{beam} \sim 1$) of the neutron rich $^{124}$Sn projectile. The opposite is true for the proton rich isotope,

$^7$Be, whose yield is larger near the original rapidity (y/y$_{beam}$ ~ 0) of the proton rich $^{112}$Sn target. For both fragments, the $\frac{dM}{dyd(p_\perp/A)}$ distributions are somewhat larger at mid-rapidity than at beam or target rapidity where the Coulomb repulsion between the fragments and the residue suppresses the fragment yield in a characteristic circular pattern about the rapidities of projectile- and target-like residues.

One advantage of choosing isobars to study the isospin dependence of fragment formation is that the collective contribution to the energy is the same for both isobars; thus the relative admixture of thermal and collective energy contributions is the same for both. However, since the isobars have different charge, contributions from the Coulomb effect could be very large. The Coulomb effect is further amplified if the yield ratios of $^7$Li and $^7$Be are plotted as in the third panel of Figure 4. This occurs because the Coulomb "circle" is larger for $^7$Be than for $^7$Li.

To disentangle the isospin effects from the Coulomb effects even in the mid-rapidity region, we construct the isospin transport ratio using Eq. 1. The non-diffusion effects of the Coulomb interaction should cancel to the first order. The bottom panel of Figure 4 shows the 2D representation of the isospin transport ratios, $R_7$=$R_i$(ln(Y($^7$Li)/Y($^7$Be)). By "normalizing" the observable ln[Y($^7$Li)/Y($^7$Be)] from the mixed systems with the two symmetric reactions, $R_7$ maximizes the effect of isospin diffusion. Comparing the bottom two panels, it is clear that effects from Coulomb force are largely cancelled, even though it is difficult to be quantitative about cancellation in the regions, y/y$_{beam}$~1 and y/y$_{beam}$~0, at small transverse momenta where the Coulomb suppression makes the isotopic yields very small. The nearly complete elimination of the Coulomb effect demonstrates the power of the isotope transport ratio and suggests that other non-diffusion effects such as pre-equilibrium emissions are most likely largely cancelled as well.

It is clear that $R_7$ makes a rapid transition through zero at mid-rapidity. It is nearly uniformly positive (red) around the neutron rich projectile rapidity and nearly uniformly negative (blue) around the neutron deficient target rapidity. The dependence on momentum at a fixed rapidity is rather weak and the dependence on the rapidity can be examined quantitatively by integrating the transport isospin ratios, $R_7$, over transverse momentum. The rapidity dependence is shown by the data points in Figure 4. $R_7$ is nearly flat near the projectile and the target rapidity regions before dropping to zero around mid-rapidity. As can be seen from the comparison of the data points with the shaded boxes that depict the earlier isoscaling analyses, both analyses provide isospin transport ratios of about 0.5±0.15 at $y/y_{beam}$ = 0.7-1.0 and -0.5±0.18 at $y/y_{beam}$=0-0.3 when the same impact parameter and rapidity gates are applied to the same systems [6]. This shows that the mirror nuclei ratios can be reliably used to assess the transport of isospin asymmetry throughout the collision.

Figure 6 shows the rapidity dependence of $R_7$ for $y/y_{beam}$ > 0.5 with three impact parameter gates: the peripheral gate ($b/b_{max}$ > 0.8) discussed earlier, a mid-impact parameter gate (0.4 < $b/b_{max}$ < 0.6) and a central collision gate (0 < $b/b_{max}$ < 0.2). To reduce the error bars, we have taken advantage of the approximate symmetry of $R_7$, i.e. $R_7(y/y_{beam})$ = $-R_7(0.5- y/y_{beam})$ to combine data from $^{124}$Sn+$^{112}$Sn collisions with data from $^{112}$Sn+$^{124}$Sn collisions. The shape of $R_7$ changes significantly with impact parameter. For $b/b_{max}$ > 0.8, $R_7$ makes a rapid transition through zero near $y/y_{beam}$ = 0.5 and then remains roughly constant at $y/y_{beam}$ > 0.7. For mid-central collisions, 0.4 < $b/b_{max}$ < 0.6, $R_7$ changes linearly with rapidity. For central collisions, 0 < $b/b_{max}$ < 0.2, $R_7$ remains close to 0 for 0.5 < $y/y_{beam}$ < 0.8 and then increases rapidly near projectile rapidity. At mid-rapidity and at projectile rapidity, within experimental uncertainties, $R_7$ are nearly equal and seem to be independent of the impact parameter. $R_7$ decreases with impact parameters at rapidity between the mid-rapidity and projectile rapidity.

It appears that the isospin asymmetry in peripheral collisions makes a rapid transition between a value characteristic of target rapidity to another value characteristic of projectile rapidity. In central collisions, there is a broad domain near $y/y_{beam}$ = 0.5 where the values for $R_7$ ($R_7 \sim 0$) suggest isospin equilibrium and then a rapid transition to values ($R_7 \sim 0.5$) similar to that characteristic of peripheral collisions for $y/y_{beam} \approx 1.0$. Such a trend is consistent with the picture that participant nucleons near $y/y_{beam}$ = 0.5 suffered collisions that mixed nucleons from the target and projectile nuclei nearly completely together, while mixing of nucleons near $y/y_{beam}$ = 1.0 is similar to that observed in peripheral collisions. We note that Dempsey et al. [20] analyzed ratios of elemental and isotopic yields and observed significant non-equilibrium effects at both central and beam rapidities for $^{136,124}$Xe+$^{112,124}$Sn collisions at E/A=55 MeV. We are not able to compare directly those results to ours because the experimental observables and analyses techniques differ significantly. Additional measurements would be useful to establish more precisely the degree to which isospin equilibrium is achieved in central collisions.

The actual trajectories of nucleons that emerge near y/y$_{beam}$ = 1.0 in central collisions is difficult to establish with the current data alone, but maybe addressed by detailed comparison with transport theory. Such comparisons lie outside the scope of the present paper. One would like to address how many of these nucleons penetrate through the target and how many have trajectories that avoid the target. Similar trends are observed for the $^{11}$C/$^{11}$B mirror nuclei ratios, albeit with significantly reduced statistics. Therefore it maybe useful to examine the emission pattern for $^{12}$C fragments. Figure 7 shows the $\frac{dM}{dyd(p_\perp/A)}$ for $^{12}$C fragments in central collisions. The red circle corresponds to constant kinetic energy in the cm frame. The measured momentum distributions are not spherical but are somewhat elongated along the beam axis. Since there is no projectile-like residues for these central collisions as shown in Figure 2, this elongation is not caused by a projectile-like residue. It suggests that transparency could occur if the mean free path is comparable to the size of the system. If so, the present data may provide useful information about isospin dependent mean free path for nucleons in these collisions.

In summary, we have demonstrated that the ratios of the yields of mirror nuclei, such as Y($^7$Li)/Y($^7$Be), can be an effective observable for studying the isospin diffusion phenomenon. Such ratios have the virtue of being less sensitive to the unknown admixtures of collective and thermal motion. We also demonstrated that the isospin asymmetry transport ratios remove the distortions due to effects such as Coulomb repulsion that could obscure the effects of the density dependence of the symmetry energy. Similar trends are observed for R$_{11}$ constructed from the Y($^{11}$B)/Y($^{11}$C) ratio. (The other mirror nuclei pair constructed from Y($^3$H)/Y($^3$He) does not appear to be as suitable for the diffusion studies because the pre-equilibrium contributions and time dependent emission effects are much larger for these light nuclei causing asymmetry between particles emitted from projectile-like and target-like rapidities.) Similarly, our study suggests that other probes involving isotope and isotone ratios in systems where isoscaling is strictly observed. When this is done, observables that should depend linearly on the isospin asymmetry δ can be defined and the isospin asymmetry transport ratios corresponding to these observables can be expected to be equal to those constructed from the isospin asymmetry δ of the source at the time when the particles are emitted.

We have explored the impact parameter and rapidity dependence of the isospin transport ratios, and of the local isospin asymmetry itself. We find that the isospin transport ratios for $^{112,124}$Sn+$^{112,124}$Sn collisions increase monotonically with rapidity. The transition from negative ratios at negative rapidity to positive ratios at positive rapidity changes with impact parameter. For large impact parameters, the change at mid-rapidity is abrupt and there is a large region of roughly constant isospin asymmetry ratio near beam and target rapidity. As the impact parameter decreases, this transition become more gradual and for central collisions, there is a large region around mid-rapidity with small isospin-transport ratios, consistent with isospin equilibrium.

This work is supported by the National Science Foundation under Grant Nos. PHY-0245009 (MSU) and by the DOE under grant numbers DE-FG02-87ER-40316 (WU) and DE-FG02-88ER-40404 (IU).

**REFERENCES**

1. P. Danielewicz, R. Lacey and W.G. Lynch, *Science* 298, 1592 (2002) and references therein.

2. J.M. Lattimer and M. Prakash, *Ap. J.* **550**, 426 (2001).

3. B. Alex Brown, *Phys. Rev. Lett.* **85**, 5296 (2000).

4. C. J. Horowitz, S. J. Pollock, P. A. Souder, R. Michaels, *Phys. Rev. C* **63**, 025501 (2001).

5. Bao-An Li, *Phys. Rev. Lett.* **85**, (2000) 4221.

6. M. B. Tsang, T. X. Liu, L. Shi, P. Danielewicz, C. K. Gelbke, X. D. Liu, W. G. Lynch, W. P. Tan, G. Verde, A. Wagner, H. S. Xu, W. A. Friedman, L. Beaulieu, B. Davin, R. T. de Souza, Y. Larochelle, T. Lefort, R. Yanez, V. E. Viola, Jr., R. J. Charity and L. G. Sobotka, *Phys. Rev. Lett.* **92**, 062701 (2004).

7. L. Shi and P. Danielewicz, *Phys. Rev. C* **68**, 064604 (2003).

8. Lie-Wen Chen, Che Ming Ko and Bao-An Li, *Phys. Rev. Lett. 94*, 032701 (2005).

9. F. Rami, Y. Leifels, B. de Schauenburg, A. Gobbi, B. Hong, J. P. Alard, A. Andronic, R. Averbeck, V. Barret, Z. Basrak, N. Bastid, I. Belyaev, A. Bendarag, G. Berek, R. aplar, N. Cindro, P. Crochet, A. Devismes, P. Dupieux, M. Delalija, M. Eskef, C. Finck, Z. Fodor, H. Folger, L. Fraysse, A. Genoux-Lubain, Y. Grigorian, Y. Grishkin, N. Herrmann, K. D. Hildenbrand, J. Kecskemeti, Y. J. Kim, P. Koczon, M. Kirejczyk, M. Korolija, R. Kotte, M. Kowalczyk, T. Kress, R. Kutsche, A. Lebedev, K. S. Lee, V. Manko, H. Merlitz, S. Mohren, D. Moisa, J. Mösner, W. Neubert, A. Nianine, D. Pelte, M. Petrovici, C. Pinkenburg, C. Plettner, W. Reisdorf, J. Ritman, D. Schüll, Z. Seres, B. Sikora, K. S. Sim, V. Simion, K. Siwek-Wilczyska, A. Somov, M. R. Stockmeier, G. Stoicea, M. Vasiliev, P. Wagner, K. Winiewski, D. Wohlfarth, J. T. Yang, I. Yushmanov, and A. Zhilin, *Phys. Rev. Lett.* **84**, 1120 (2000).

10. M. B. Tsang, C. K. Gelbke, X. D. Liu, W. G. Lynch, W. P. Tan, G. Verde, H. S. Xu, W. A. Friedman, R. Donangelo, S. R. Souza, C. B. Das, S. Das Gupta, and D. Zhabinsky, *Phys. Rev. C* **64**, 054615 (2001).

11. A. S. Botvina, O. V. Lozhkin, and W. Trautmann, *Phys. Rev. C* **65**, 044610 (2002).

12. Akira Ono, P. Danielewicz, W. A. Friedman, W. G. Lynch, and M. B. Tsang, *Phys. Rev. C* **68**, 051601(R) (2003).

13. T. X. Liu, M. J. van Goethem, X. D. Liu, W. G. Lynch, R. Shomin, W. P. Tan, M. B. Tsang, G. Verde, A. Wagner, H. F. Xi, and H. S. Xu, M. Colonna and M. Di Toro, M. Zielinska-Pfabe, H. H. Wolter, L. Beaulieu, B. Davin, Y. Larochelle, T. Lefort, R. T. de Souza, R. Yanez, and V. E. Viola, R. J. Charity and L. G. Sobotka, *Phys. Rev. C* **69**, 014603 (2004).

14. H. S. Xu, M. B. Tsang, T. X. Liu, X. D. Liu, W. G. Lynch, W. P. Tan, A. Vander Molen, G. Verde, A. Wagner, H. F. Xi, C. K. Gelbke, L. Beaulieu, B. Davin, Y. Larochelle, T. Lefort, R. T. de Souza, R. Yanez, V. E. Viola, R. J. Charity and L. G. Sobotka, *Phys. Rev. Lett.* **85**, 716 (2000).

15. T. X. Liu, W. G. Lynch, M. J. van Goethem, X. D. Liu, R. Shomin, W. P. Tan, M. B. Tsang, G. Verde, A. Wagner, H. F. Xi, and H. S. Xu, W.A. Friedman, S.R. Souza, R. Donangelo, L. Beaulieu, B. Davin, Y. Larochelle, T. Lefort, R. T. de Souza, R. Yanez, and V. E. Viola, R. J. Charity and L. G. Sobotka, *Eruophys. Lett. 74, 806 (2006)*

16. R.T. DeSouza, N. Carlin, Y.D. Kim, J. Ottarson, L. Phair, D.R. Bowman, C.K. Gelbke, W.G. Gong, W.G. Lynch, R.A. Pelak, T. Peterson, G. Poggi, M.B. Tsang, and H.M. Xu, *Nucl. Inst. and Meth*. A **295**, 109 (1990).

17. A. Wagner, W.P. Tan, K. Chalut, R.J. Charity, B. Davin, Y. Larochelle, M.D. Lennek, T.X. Liu, X.D. Liu, W.G. Lynch, A.M. Ramos, R. Shomin, L.G. Sobotka, R.T. De Souza, M.B. Tsang, G. Verde and H.S. Xu, Nucl. Instr. Meth. Phys. Res. A 456, 290 (2001).

18. B. Davin, R.T. de Souza, R. Yanez, Y. Larochelle, R. Alfaro, H.S. Xu, A. Alexander, K. Bastin, L. Beaulieu, J. Dorsett, G. Fleener, L. Gelovani, T. Lefort, J. Poehhman, R.J. Charity, L.G. Sobotka, J. Elson, A. Wagner, T.X. Liu, X.D. Liu, W.G. Lynch, L. Morris,

R. Shomin, W.P. Tan, M.B. Tsang, G. Verde, and J.Yurkon, *Nucl. Inst. and Meth A* **473**, 302 (2001).

19. H.S. Xu, M.B. Tsang, T.X. Liu, X.D. Liu, W.G. Lynch, W.P.Tan and G. Verde, L. Beaulieu, B. Davin, Y. Larochelle, T. Lefort, R.T. de Souza, R. Yanez, V.E. Viola, R.J. Charity, and L.G. Sobotka, Phys. Rev. Lett. **85**, 716 (2000).

20. J. F. Dempsey, R.J. Charity, L.G. Sobotka, G.J. Kunde, S. Gaff, C.K. Gelbke, T. Glasmacher, M.J. Huang, R. Lemmon, W.G. Lynch, L. Manduci, L. Martin, M.B. Tsang, J.D. Agnihotri, B. Djerroud, W.U. Schroder, W. Skulski, J. Toke, Phys. Rev. **C54**, 1710 (1996).

**Figure Captions:**

Figure 1: Upper panel: The solid circles, solid squares and open diamonds show the experimental data for α, β and α-β, respectively, that have been measured for central Sn+Sn collisions. The x axis corresponds to the isospin asymmetry of the combined system. Lower panel: The solid circles, and solid squares show the experimental data for $X_7$ and $X_{11}$, respectively, that has been measured for central Sn+Sn collisions.

Figure 2: Correlation between the charge of the projectile-like fragment and observed total charge particle multiplicity. The corresponding values of $b/b_{max}$ are given in the upper axis of the figure. The numbers indicating the contour levels are in arbitrary units.

Figure 3: The solid circles, solid stars, open stars and open circles show the experimental isotopic distributions at $y/y_{beam} > 0.7$ and $b/b_{max} > 0.8$ that has been measured for $^{124}Sn+^{124}Sn$, $^{124}Sn+^{112}Sn$, $^{112}Sn+^{124}Sn$, and $^{112}Sn+^{112}Sn$ collisions, respectively. The solid lines are drawn to guide the eyes and the dashed lines are discussed in the text.

Figure 4: The solid rectangles at $0 < y/y_{beam} < 0.3$ and $0.7 < y/y_{beam} < 1.0$ show the values of $R_i(\alpha)$ measured for the $^{124}Sn+^{112}Sn$ and $^{112}Sn+^{124}Sn$ systems, respectively [6]. The data points show the values of $R_i(X_7)$ measured for A=7 mirror nuclei in peripheral $^{124}Sn+^{112}Sn$ collisions as a function of rapidity.

Figure 5 $\frac{dM}{dyd(p_\perp/A)}$ distribution as function of rapidity, $y/y_{beam}$, measured in $^{124}Sn+^{112}Sn$ collisions at E/A=50 MeV. From the top, first panel: $\frac{dM}{dyd(p_\perp/A)}$ distribution for $^7Li$ fragments. Second panel: $\frac{dM}{dyd(p_\perp/A)}$ distribution for $^7Be$ fragments. Third Panel: Ratio of the $\frac{dM}{dyd(p_\perp/A)}$ distribution for $^7Li$ fragments divided by the $\frac{dM}{dyd(p_\perp/A)}$ distribution for $^7Be$. Fourth panel: $R_7$ distribution.

Figure 6: The open diamonds, solid squares and solid circles show the rapidity dependencies of $R_7$ for central, mid-central and mid-peripheral collisions, respectively.

Figure 7: $\frac{dM}{dyd(p_\perp/A)}$ distribution for 12C fragments produced in central $^{124}Sn+^{112}Sn$ collisions at E/A=50 MeV. The dashed lines show the boundaries at $\theta_{cm} = 70°$ and $110°$ for the mid-rapidity gate that was used to create the data shown in Figure 1 [13]. The circle corresponds to the contour of constant velocity in the center of mass.

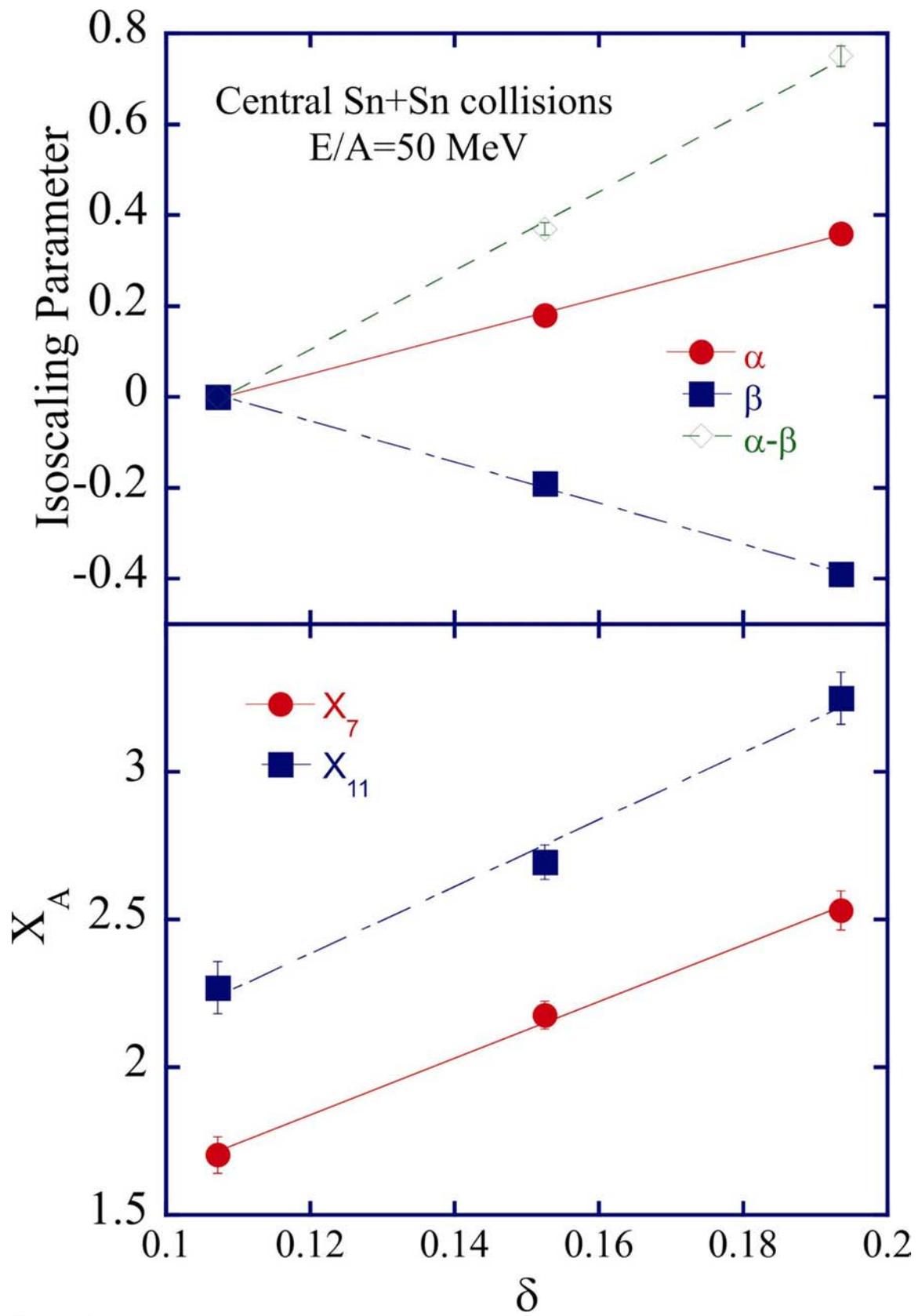

Figure 1

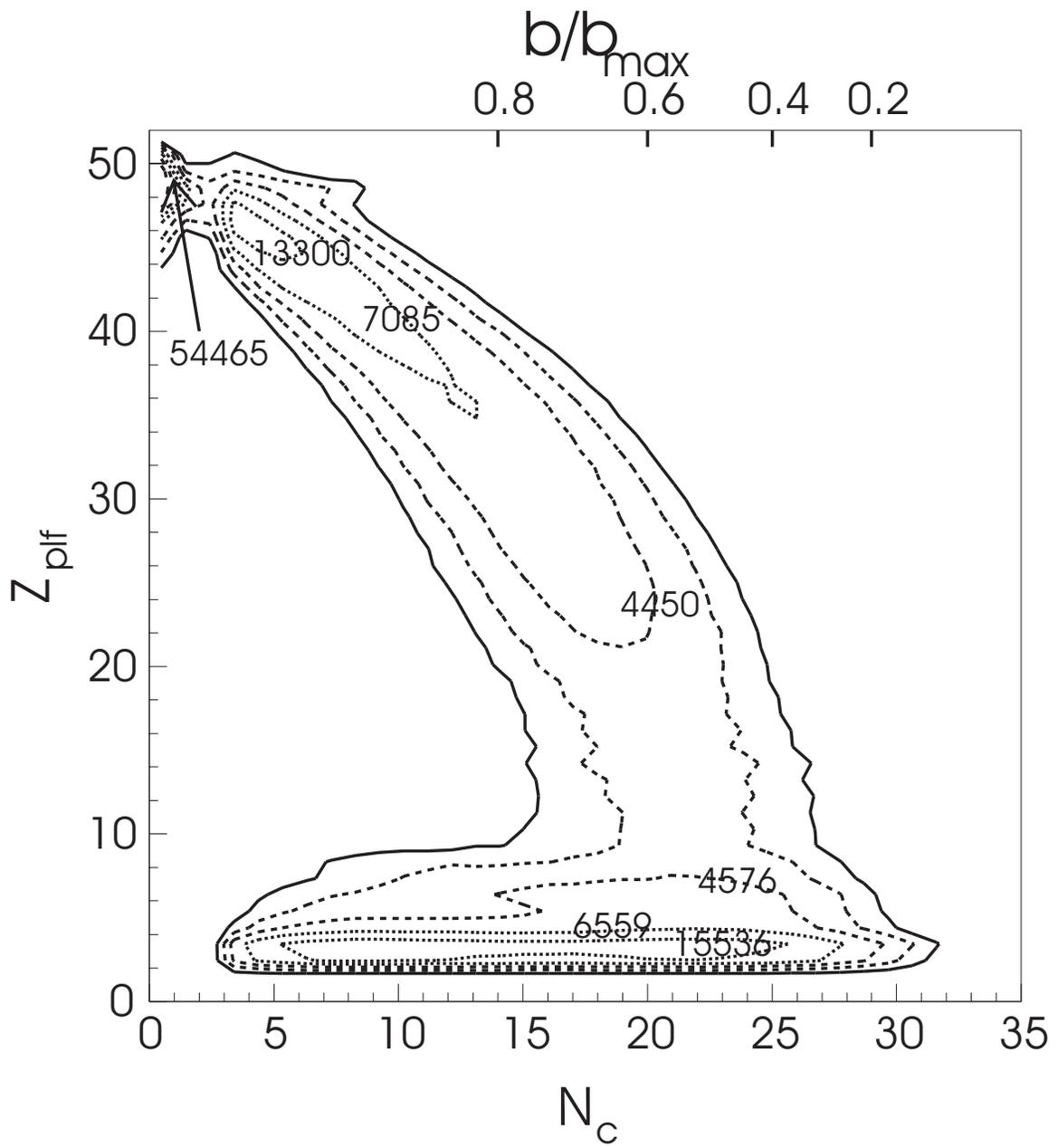

Figure 2

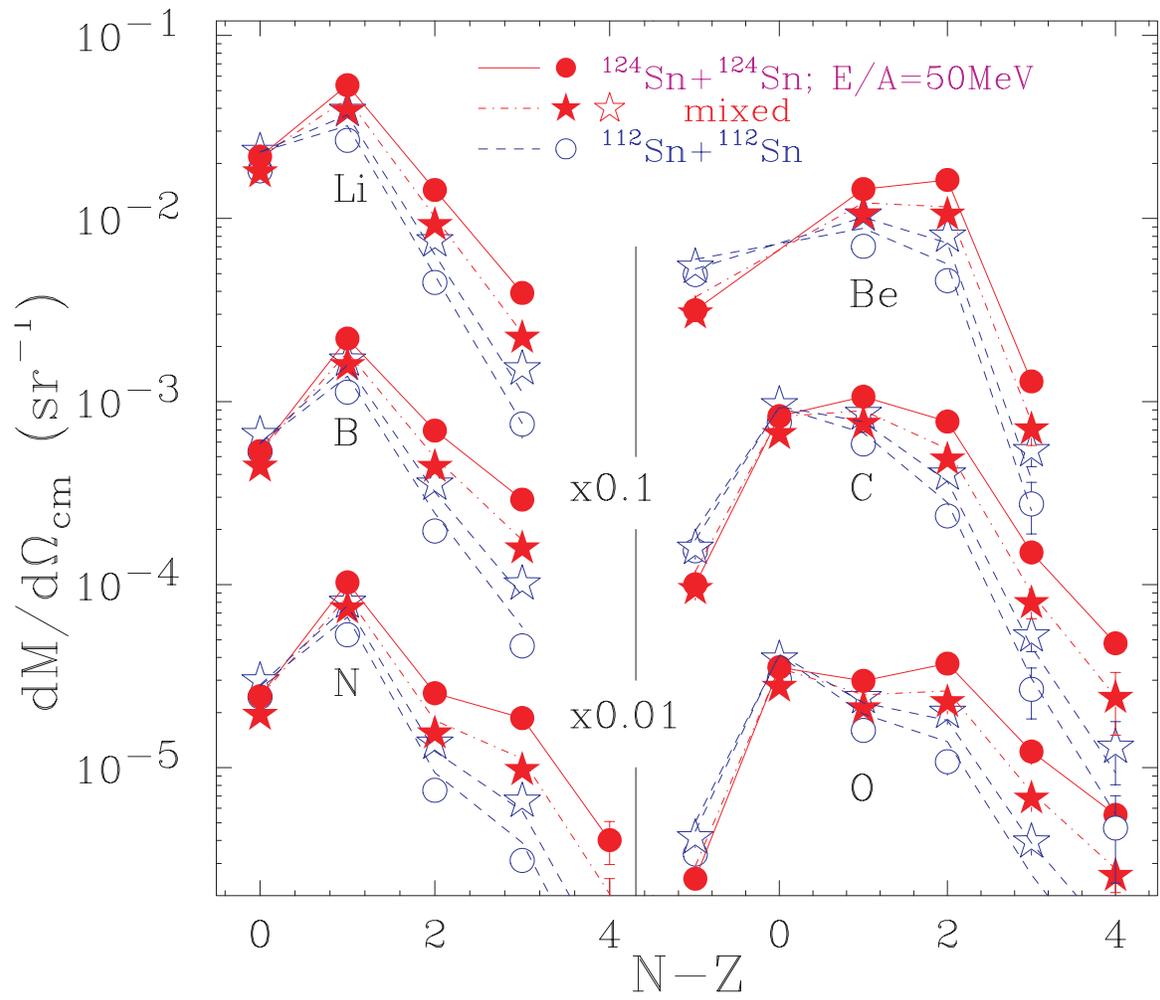

Figure 3

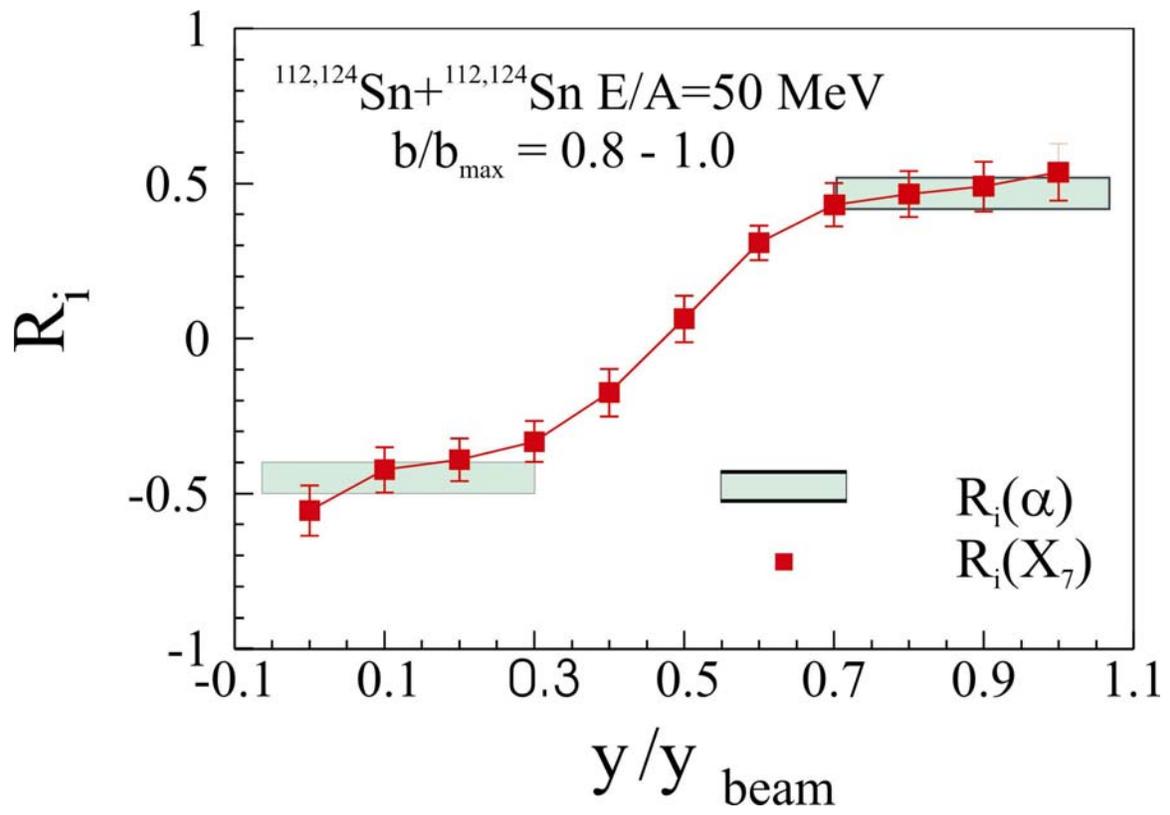

Figure 4

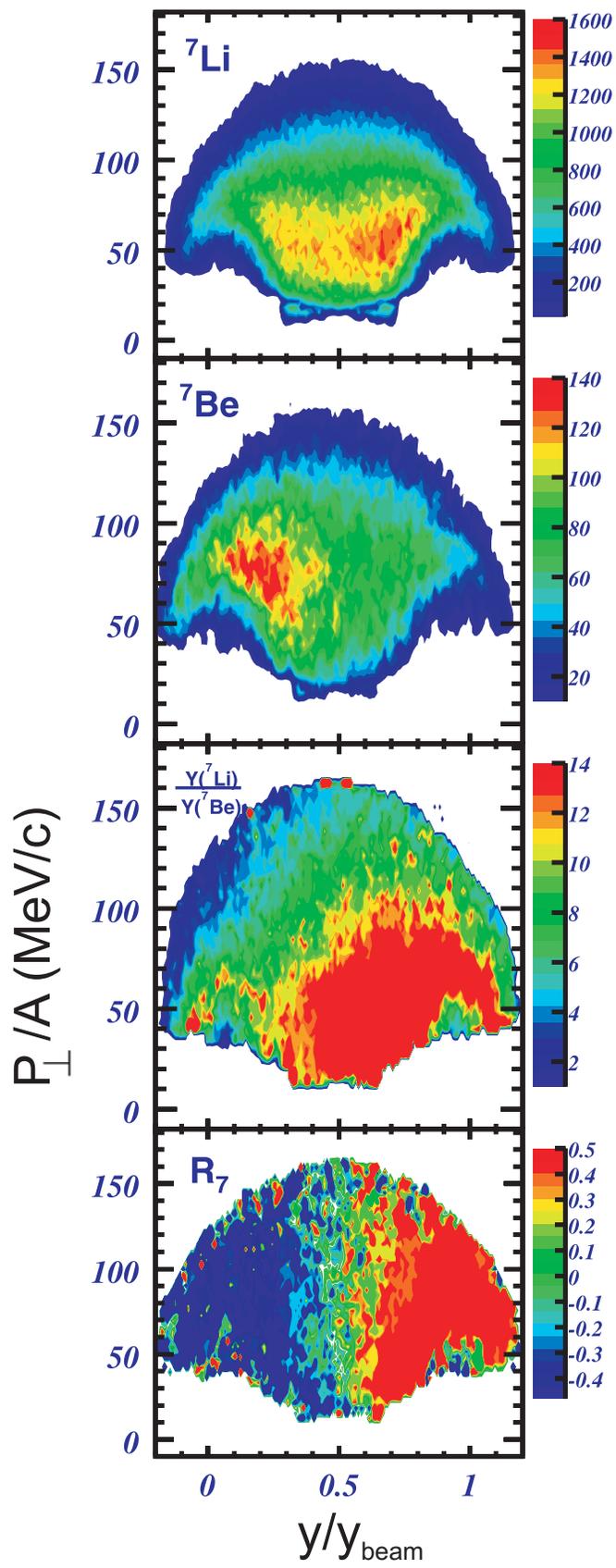

Figure 5

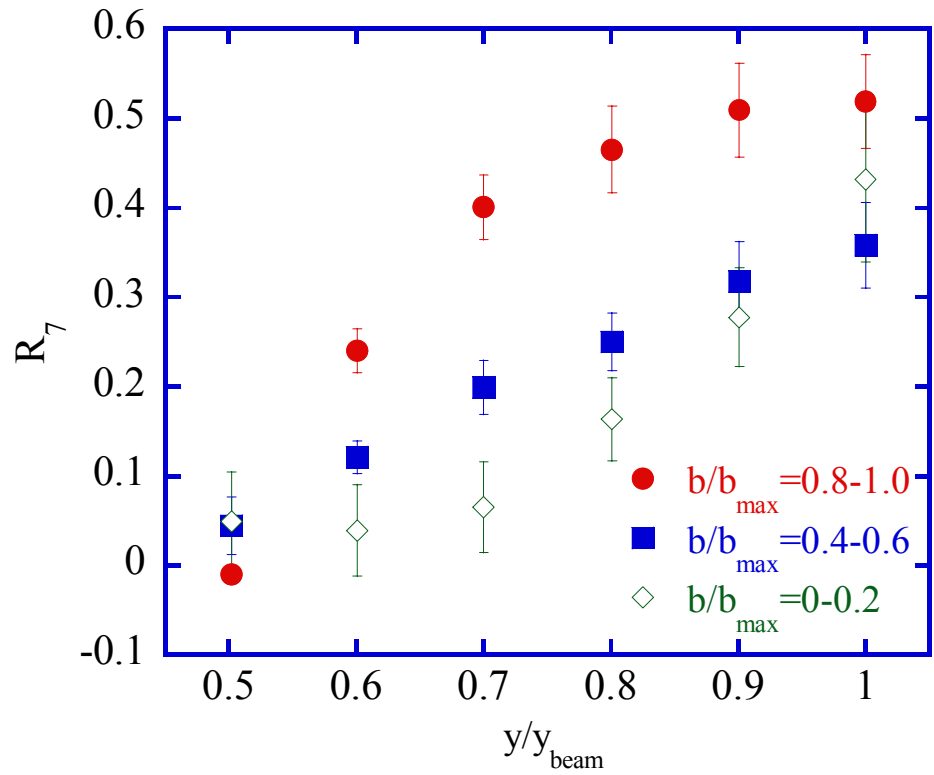

Figure 6

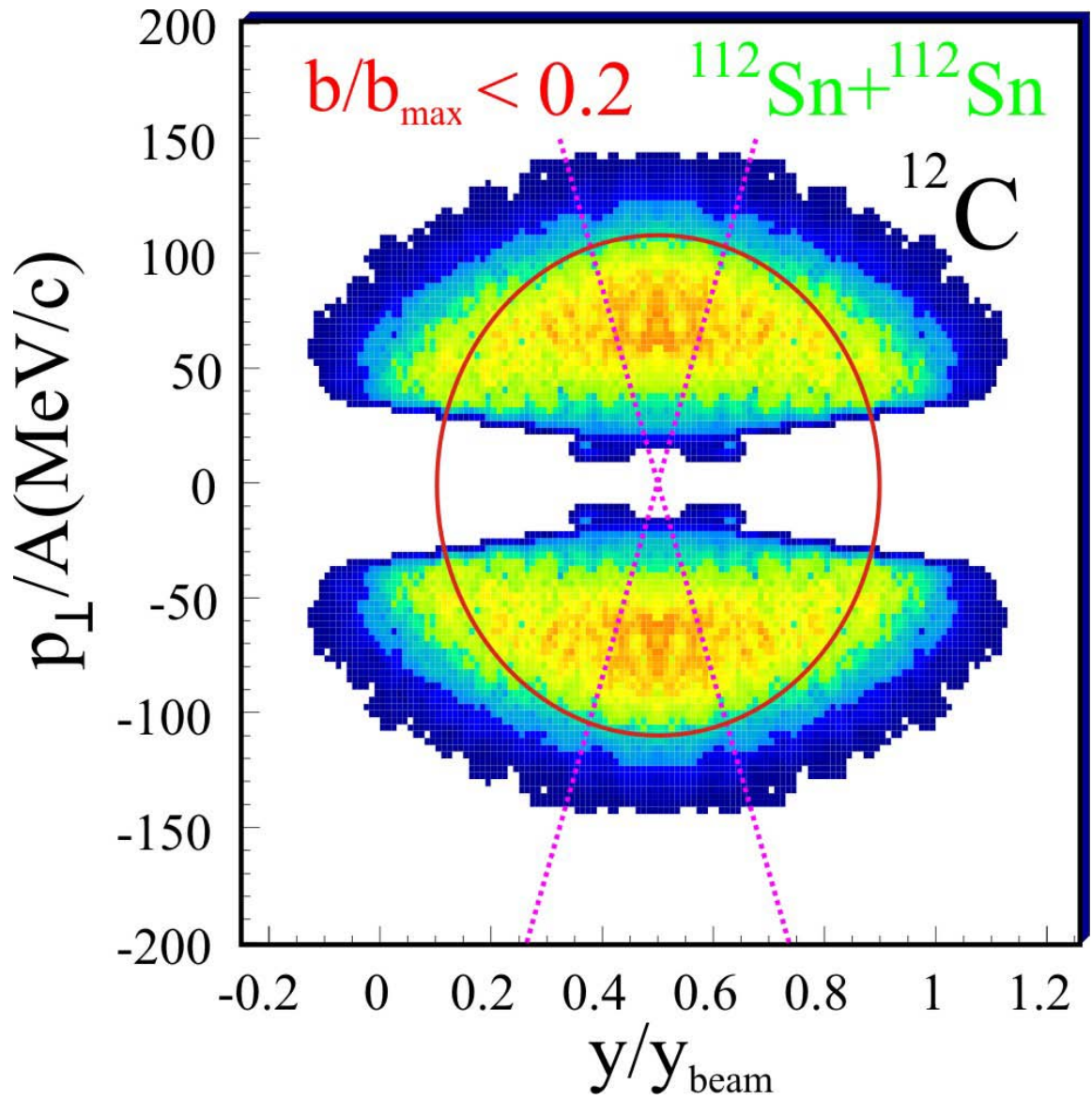

Figure 7